\documentclass[journal]{IEEEtran}
\ifCLASSINFOpdf
\else
\fi

\usepackage{graphicx, lineno, float, amsmath, enumitem, booktabs, caption, multirow}
\usepackage{algorithm}
\usepackage{algorithmic}
\usepackage{multirow}
\usepackage{threeparttable}
\usepackage{booktabs}
\usepackage{tabularx}
\usepackage{colortbl}

\begin{document}
	\title{Multi-Representation Knowledge Distillation For Audio Classification}

	\author{Liang Gao,~\IEEEmembership{}
		Kele Xu,~\IEEEmembership{}
		Huaimin Wang,~\IEEEmembership{}
		Yuxing Peng~\IEEEmembership{}
		%Huaimin Wang.~\IEEEmembership{}% <-this % stops a space
		\thanks{L. Gao, K. Xu, H. Wang and Y. Peng were with the National Key Laboratory of Parallel and  Distributed Processing, College of Computer, National University of Defense Technology, Changsha 410073, China e-mail: (Kelele.xu@gmail.com).}% <-this % stops a space
	}


	\markboth{}%
	{Shell \MakeLowercase{\textit{et al.}}: Bare Demo of IEEEtran.cls for IEEE Journals}



	\maketitle
	
	\begin{abstract}
		As an important component of multimedia analysis tasks, audio classification aims to discriminate between different audio signal types and has received intensive attention due to its wide applications. Generally speaking, the raw signal can be transformed into various representations (such as Short Time Fourier Transform and Mel Frequency Cepstral Coefficients), and information implied in different representations can be complementary. Ensembling the models trained on different representations can greatly boost the classification performance, however, making inference using a large number of models is cumbersome and computationally expensive. In this paper, we propose a novel end-to-end collaborative learning framework for the audio classification task. The framework takes multiple representations as the input to train the models in parallel. The complementary information provided by different representations is shared by knowledge distillation. Consequently, the performance of each model can be significantly promoted without increasing the computational overhead in the inference stage. Extensive experimental results demonstrate that the proposed approach can improve the classification performance and achieve state-of-the-art results on both acoustic scene classification tasks and general audio tagging tasks.
	\end{abstract}
	
	\begin{IEEEkeywords}
		convolutional neural networks, acoustic classification, knowledge distillation.
	\end{IEEEkeywords}

	% \fi
	\IEEEpeerreviewmaketitle

	\section{Introduction}
	\IEEEPARstart{A}{udio} classification task refers to identify a pre-defined label for an audio signal \cite{virtanen2018computational}. The potential applications of audio classification seem to be evident in several fields, such as multimedia retrieval, security surveillance \cite{clavel2005events}, health care monitoring \cite{peng2009healthcare} and context-aware services \cite{ma2003context}. Due to the dramatic increase of the sound recordings, the demand for automatic audio classification is growing rapidly in last decades. Sustainable efforts have been made to address the audio classification problems \cite{virtanen2018computational,xu2018large, dhanalakshmi2009classification,lee2006automatic,dhanalakshmi2011classification,sainath2015learning,choi2017convolutional}.
	
	The launch of the Detection and Classification of Acoustic Scenes and Events (DCASE) \cite{7100934} challenge promoted the development of audio classification, which was organized by the IEEE Audio and Acoustic Signal Processing (AASP) technical committee. Many audio processing techniques have been proposed during the past years, and the applications of deep learning in the audio classification have witnessed a significant increase, especially the convolutional neural network (CNN). The traditional method, which commonly involve models like Gaussian Mixture Model (GMM) \cite{dhanalakshmi2011classification}, Support Vector Machine (SVM) \cite{dhanalakshmi2009classification} or Hidden Markov Model (HMM) \cite{lee2006automatic} are trained using the frame-level features such as Mel-frequency Cepstral Coefficients (MFCC) \cite{mesaros2018detection}.
	
	\cite{lee2018samplecnn,lee2017sample-level} used one-dimensional convolution and fully connected layers to learn from the original raw signal; \cite {piczak2015environmental} combined MFCC and its delta into two-channel data, and used convolutional neural networks for feature extraction and classifier training. \cite{8085174} extracted three channels of log Mel-spectrograms (static, delta, and delta delta) as the DCNN input for speech emotion recognition. \cite {sercu2016dense} explored the use of dilated convolutions to use more contextual information to classify audio. \cite{sainath2015convolutional,sainath2015learning,hoshen2015speech} applied the time-domain convolution method to different tasks, such as speech recognition and sound event detection. \cite{choi2017convolutional} used convolutional recurrent neural network (CRNN) for music labeling tasks.

	% The problems of Previous attempts
	In most existing audio classification works \cite{piczak2015environmental,sercu2016dense,sainath2015convolutional,sainath2015learning,choi2017convolutional,zhang2019large,wang2019acoustic}, the raw signal was transformed into one single representation (for example, Short Time Fourier Transform (STFT) \cite{ramalingam2006gaussian} logMel and MFCC), then train the classifiers based on the single representation. The performance of deep neural networks heavily relied on the representation of the audio clip while one single representation may cannot present the information effectively and efficiently. The conversion process from the original audio signal to the advanced representation undergone a variety of transformations and information compression operations, which undoubtedly led to the loss of audio information. Thus, single representation-based deep models are still short of accuracy. Fusion the knowledge obtained using different representations can greatly improve the classification performance \cite{yin2018learning, xu2019general}, as the single representation may be stuck at poor local minimums during the training phase.
	
	Different representations represent different aspects of signals, joint using the knowledge of different representations could enhance model generalization performance \cite{xu2018meta}. For example, in image classification, \cite{8059841} combined the discriminative power of different views to jointly learn the classifiers and transformation matrices. There are many works that improve model performance by ensemble networks \cite{dietterich2000ensemble,rokach2010ensemble,wang2018weakly} which trained on different representations. But it leads to the increase of model complexity, some researchers tried to fusion information of different representations into a single network. In audio classification, \cite{8490588} combines convolutional neural networks (CNNs) with long short-term memory (LSTM) to exploit the correlative information from multiple views. These methods require careful model design, The universal approach to employ complementary information provided by different representations is still under-explored.	
	
	Intuitively, there are two kinds of approaches for the utilization of complementary information from multiple representations. One straightforward way is early-fusion, which concatenates different representations as a whole input to a single network, with each representation being a separate channel. However, this method decreases the network performance in the practical settings. Another method is later-fusion \cite{yin2018learning,dietterich2000ensemble}, which ensembles the predictions generated by different classifiers, which can empirically boost the classification performance. Nevertheless, the inference of a large number of models is cumbersome and computationally expensive.

	Recently, it has been found that knowledge distillation \cite{hinton2015distilling,zhang2018deep} can be used to transfer knowledge between different models, which could improve the classification without increasing the computational complexity in the inference phase \cite{zhang2018deep,sun2017ensemble}. Inspired by the knowledge distillation, in this paper, a multi-representation based knowledge distillation approach was proposed for audio classification, with the goal to fully utilize complementary information introduced by different representations of the audios. Moreover, our method uses only one model in the inference phase, so the computational cost is independent of the number of models that participated in the distillation framework. Overall, our contributions are three-fold:
	
	\begin{itemize}
	\item Firstly, a novel collaborative learning framework is proposed for the audio classification task. Unlike most of the traditional approaches which only use a single representation, we leverage multiple representations within the framework. Complementary information embedded in multiple representations is extracted through different neural networks and fused in the end-to-end distillation framework. The fused knowledge is then fed back to each network during the training phase and can effectively improve the performance of different models. Consequently, the performance of each model has been improved by using collaborative distillation in the training stage. Moreover, different network architectures can be easily integrated into the framework.
	
	\item Secondly, our method provides a novel ensemble approach without additional inferring cost. Due to the decoupling nature of the knowledge distillation framework, no dependency is enforced between participated models. In other words, each model can be used independently in the inference stage. Those lightweight models trained by distillation become very effective in resource-constrained computing scenarios.
	
	\item Thirdly, extensive experiments are conducted on acoustic scene classification (DCASE 2018 Challenge Task 1A) and general audio tagging (DCASE 2018 Challenge Task 2). We find that the learning framework can improve the performance of audio classification and achieve state-of-the-art results on both acoustic scene classification and general audio tagging task. More specifically, with our framework single network obtained the mAP@3 of 93.26\% in the acoustic scene classification task, and the accuracy of 72.48\% in the acoustic scene classification task.
	
	\end{itemize}

	The paper is organized as follows. We firstly discuss the relationship between our method and prior works in section \ref{related}, while the details of the proposed approach are presented in section \ref{method}. The experimental settings, the analysis of results and conclusions are given in the last three parts.

	\section{Related Work}
	\label{related}
	In this section, we discuss the relationship between our work and previous work, which includes two parts: audio classification and knowledge distillation.
	
	\subsection{Audio classification}
	Audio tagging aims to predict the presence or absence of certain acoustic events in the interested acoustic scene. Some traditional methods for audio classification like SVM, HMM and GMM are applied in industry. In the first edition of DCASE in 2013, SVM \cite{burges1998tutorial,phan2016label} and bagging of decision trees \cite{dietterich2000experimental} were used. Recently, deep neural networks have shown improved performance for the audio classification task. In brief, the main modifications of current deep learning-based audio classification can be divided into four types: jointly using different representations of the audio signal \cite{kulkarni2009audio,aytar2016soundnet}; more sophisticated deep learning architectures \cite{huang2017densely,he2016deep,zhu2018environmental}; and the applications of different regularization methods (such as data augmentation) \cite{poggio1990regularization,xu2018mixup,wei2018sample,feng2018sample}.
	
	Among all of the studies, the selection of representation for the audio signal is one of the key factors for classification performance, while only a few attempts have been made in previous studies \cite{poggio1990regularization}. The audio signal can be transformed into various representations, such as raw wave signal, MFCC \cite{aucouturier2007bag}, i-vector \cite{dehak2010front} and so on. A suitable representation can effectively improve the generalization ability of the model, MFCC and logMel have been proven to be useful in CNNs. \cite{7927482} presents a novel two-phase method for audio representation, they take into account both global structure and local structure, the learned representation can effectively represent the structure of audio. \cite{7592459} argue that an image-like spectrogram cannot well capture the complex texture details of the spectrogram, so that they proposed a multichannel LBP feature to improve the robustness to the audio noise.
	Combining knowledge of multiple audio representations can obtain more comprehensive features and strengthen the generalization of models. It is found that ensembling \cite{kuncheva2003measures,sollich1996learning} the predictions generated by different classifiers can greatly boost the audio classification performance. However, making inferences using a large number of models is cumbersome and computationally expensive. On the other hand, an efficient fusion of different representations within the end-to-end manner also draws lots of attention \cite{yin2018learning}. Still, the audio classification using multi-representation needs to be thoroughly explored.
	
	\subsection{Knowledge distillation}
	Knowledge Distillation was firstly proposed in \cite{bucilua2006model} and re-popularized with the goal of model compression in \cite{hinton2015distilling}. With the knowledge distillation method, knowledge of pre-trained complex models can be transferred to a small network, which would help to improve the model performance. Except for the traditional supervised learning objective such as the cross-entropy loss which based on the ground truth label, distillation hopes to introduce the extra supervision from the teacher model to the student model. The extra supervision can be in the forms of classification probabilities \cite{hinton2015distilling,gao2019multistructure} or feature representation \cite{kim2018paraphrasing}.
	
	\cite{sun2017ensemble} uses the knowledge distillation method to extract knowledge from an integrated model and compress the knowledge into a single network. In \cite{zhang2018deep}, two models are training at the same time, exchanging their prediction probability with each other to enhance the model performance.
	For classifiers that use label smoothing, soft labels replaced one-hot hard labels. Label smoothing can reduce the interference of noise data or wrong labels. In online distillation \cite{anil2018large} studied the co-distillation of multiple examples of neural networks, which using exactly the same settings, and achieved training acceleration.
	\cite{batra2017cooperative} proposed cooperative learning, in which jointly trains multiple models in different fields. For example, in the image detection task, one model inputs using the RGB image and the other model inputs using the depth image. The two models exchange the unchanged object attributes in the task so that the same task can be trained with different data inputs. Only a few attempts have been made to leverage knowledge distillation for the audio analysis tasks.

	\begin{figure*}[ht!]
		\centering
		\includegraphics[scale=0.6]{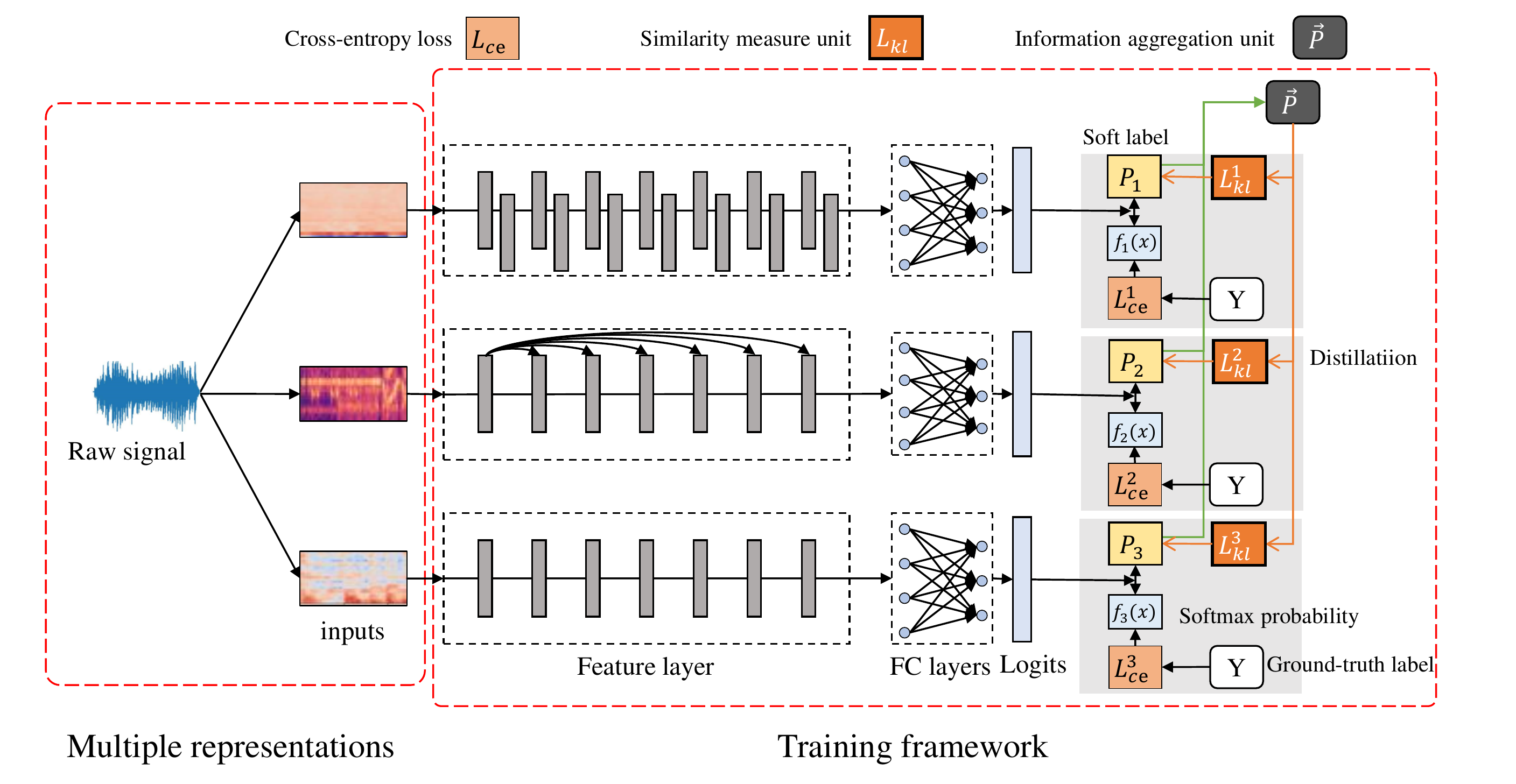}
		\caption{The overview of the proposed framework, which includes two components: (1) The multiple representations which provide input for the networks. (2) The training framework which consists of multiple branch networks (students), the information aggregation unit and the similarity measure units. In the training process, the information aggregation unit aggregate knowledge from the multiple participated networks of branches. For each of participated branches, there is a similarity measure unit that feeds back the aggregated knowledge to the network in branch.}
		\label{fig1}
	\end{figure*}

	\section{Methodology}
	\label{method}
	
	In this section, we present our approach for the audio classification task using knowledge distillation. The overview of the multi-representation distillation framework is illustrated in Figure \ref{fig1}. The framework contains multi-branch networks with leveraging multi-representation as the input. The audio signals can be transformed to different representation, which exhibits heterogeneous properties. Each representation presents a different view of the raw audio, and each view has its own individual representation space and dynamics.
	The training framework consists of multiple branches, the information aggregation unit and the similarity measure units. Each of branches in the framework is a network of full function with feature layers capturing features, fully-connected layers and logits layer for audio classification. The information aggregation unit which aggregates knowledge from multi-branch networks, and networks in the framework learn from the aggregated knowledge by minimizing the loss of similarity unit.
	The raw signal of audio can be transformed into a variety of representations with complementary knowledge. And each branch trains network with one representation. After the information aggregation unit aggregated knowledge from multi-branch networks, the branches in the framework learning from the aggregated knowledge.
	
	To train networks with the framework, first we transform the audio into different representations. And then we use the strategy of cyclic distillation for network training. Each cycle of the training process is divided into three small phases: the training of single branches, the information fusion and the distillation training.
	
	Assuming that given a training set which contains $N$ samples $X=\{x_1,x_2,..,x_N\}$, the samples come from $M$ categories, and their corresponding labels are $Y=\{y_1,y_2,...y_N\}$. Assuming there are $\Gamma$ branches in the framework.
	
	\subsection{Data preparation}
	
	In the data preparation phase, multiple representations of the audio data are generated and used as input for different branches. In acoustic classification tasks, converting the original audio data into a suitable feature representation often leads to better performance. In general, the audio signal conversion process includes framed windowing, Fourier transforming, power spectrum calculation, filtering, and discrete cosine transforming, etc.
	
	The continuous sound signal is converted into a high-level representation through time-frequency transformation, which can highlight the frequency domain features of the audio. And the transformed features representations have the advantages of low dimensions which can be represented as two-dimensional images. On the other hand, advanced audio representations are more in line with human ear characteristics.
	For example, the MFCC could simulate the human ear’s masking effect (the human ears are more sensitive to low-frequency sounds than to high-frequency sounds and more sensitive to high loudness sounds than low loudness sounds). In the transform processing of raw signal to MFCC or logMel, a set of bandpass filters which distributed gradually sparse from low to high frequencies are arranged in the critical frequency bandwidth to filter the input signal. The basic features obtained with the bandpass filters can be deployed as inputs for networks, and the generated representation is more robust and has better recognition performance in the signal with a lower noise ratio. On the other hand, the constant Q transform (CQT) avoids the disadvantage of uniform time-frequency resolution. For low-frequency waves, its bandwidth is very small, but it has higher frequency resolution to decompose similar notes; and for high-frequency waves, the bandwidth is relatively large, so that there is a higher time resolution at high frequencies to track rapidly changing overtones.
	\begin{figure}[!ht]
		\centering
		\includegraphics[scale=0.4]{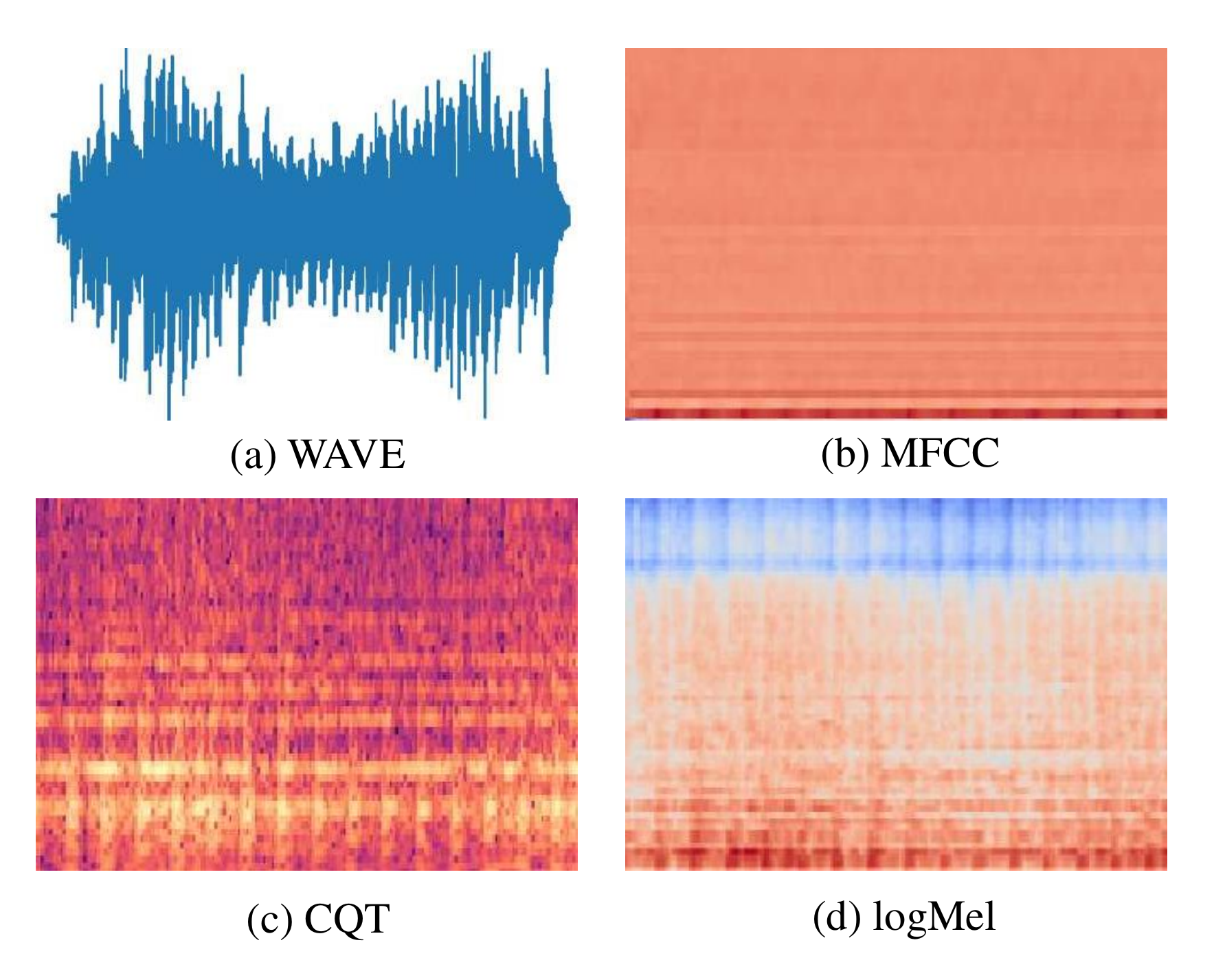}
		\caption{Different representations of audio data. (a)WAVE (b) MFCC (c) CQT and (d) logMel}
		\label{fig2}
	\end{figure}

	As shown in Figure \ref{fig2}, the audio signals could be transformed into different representations, such as log-scaled Mel-spectrograms (logMel), constant-Q transform (CQT) and MFCC and so on. Different audio representations have their own advantages, and one single representation cannot present the information effectively and efficiently. Learning to integrate multiple representations can utilize their mutual complement information. The neural networks learned with multiple representations may give better classification performance due to the complementary information contains in different representations.
	
	In the framework with $\Gamma$ branches, we convert the raw signal $D= (X, Y)$ into $\Gamma$ kinds of different representations, which are denoted as $D^i= (X^i, Y^i), i\in{(1,2,...,\Gamma)}$. 
	
	Using knowledge distillation methods to combine with networks of different structures. Each branch in the framework can independently select the network structure, assuming that the networks in i-th branch denoted as $f_i$ in the knowledge distillation framework. And $D_i$ is the corresponding training set of the classifier $f_i$.

	\subsection{Single branch training}
	In the phase of single branch training, the learning objective is to fit the ground-truth label. We minimize the cross-entropy loss between the predicted values with target labels, and the formula is as follows:
	\begin{equation}\label{equation3}
	L_{ce} = -\frac{1}{N}\sum_{i=0}^{N}[y_i\log{f(x_i)}],
	\end{equation}
	
	where $L_{ce}$ is the loss of single branch training phase for the branch network $f$.
	The cross-entropy represents the distance between the predicted value and the expected value (ground-truth label).
	
	\subsection{Information fusion}
	In order to aggregate knowledge from multiple branches, we averaged the predicted values, using soft labels as information carriers. The One-hot label only indicates the category to which the sample belongs, but ignores the similar relationship between the sample and different categories. Soft labels are the soften softmax probability of the logits layer, which consists of the sample similarity information. For each branch, we calculate the soft labels for all training data and then send them to the information aggregation unit.
	For the sample $x_i$, the formula to calculate soft label is:
	
	\begin{equation}\label{equation5}
	p_{i}= \frac{\exp(g_i/T)}{\sum_{j=0}^{M}\exp(g_j/T)},
	\end{equation}
	
	where $g_i$ is the logits layer output of a branch network $f$ corresponding to the i-th category. And $T$ is a soften hyper-parameter. The larger the value of $T$, the smoother the soft label distribution.

	The information from different branch networks is aggregated in the aggregation unit, which would be adopted as teacher information during the distillation training phase.
	The aggregated information was calculated from the multiple branches with average ensemble method:
	
	\begin{equation}\label{equation6}
	\overrightarrow{P} = \frac{1}{\Gamma}\sum_{i=1}^{\Gamma}{(P_i)}.
	\end{equation}
	
	In this equation, $P_i$ represents the training set's soft labels of branch networks $f_i$, and $\Gamma$ represents the number of branches. The averaged soft labels $\overrightarrow{P}$ has the same effect as probability of ensembled network, which is smoother and stronger generalized than the soft labels in single branch network.
	
	\subsection{Knowledge distillation}
	
	Knowledge transfer in knowledge distillation is accomplished by reducing the difference of information between teachers and students. Kullback-Leibler (KL) divergence could measure the difference of two distributions.
	In the distillation training process, the averaged soft labels which had aggregated knowledge from all branches were used as the teacher. For each branch, the similarity between soft labels $P$ and the averaged soft labels $\overrightarrow{P}$ is calculated as follows:

	\begin{equation}\label{equation7}
	L_{kl} = -\frac{1}{N}\sum_{i=0}^{N}\sum_{j=0}^M[P_{i}^j\log{\frac{\overrightarrow{P_{i}^j}}{P_{i}^j}}],
	\end{equation}

	where $P_{i}$ denote the soft labels for i-th sample of branch networks $f$, and $j$ donate the j-th category.
	
	In previous attempts for knowledge distillation \cite{zhang2018deep,lan2018knowledge}, it has been found that combining supervision from onehot labels with supervision of teacher information leads to smoother optimization and better-performed network. The distillation loss for each branch of in the distillation training phase is given as follows:
	
	\begin{equation}\label{equation9}
	L_{d} = L_{ce}+L_{kl}.
	\end{equation}

	The training process of the framework is summarized in Algorithm \ref{algorithm1}. Different from traditional two-stage distillation, we adopted cyclic distillation strategy, a large cycle including three phases: the branch training, knowledge aggregation, and distillation training. Three phases correspond to the information generation, information aggregation, and information feedback, respectively. The cyclic training could help networks better capture and utilize complementary information of multiple representations. After the model converges, any branch classifier in the framework could be applied for inferring according to the data representation or the resources limits. If better classification performance is sought, an ensemble network of multiple branch networks can also be used.

	\begin{algorithm}[t]
		\caption{\small Multi-representation knowledge distillation \label{algorithm1}}
		
		INPUT: Dataset $D=(X,Y)$,learning rate $lr$, the number of training epoch $Q$, the number of branches $\Gamma$, soften temperature parameter $T$, single branch training epoch $b$  and distillation training epoch $d$.\\
		PREPARING WORKS: Transforming the raw audio signal to multiple representations. Determining the network architecture $f_i$ and input representation $X_i$ for each branch in the framework. 
		
		\textbf{CYCLIC DISTILLATION:}
		\begin{algorithmic}[1]
			\STATE Initialization: $q=1$, load pre-train parameters $\Theta$ for branch networks.

			\FOR {$q<Q$}
			\REQUIRE {SINGLE BRANCH TRAINING:}
			\WHILE {t in range(b)}
			\STATE Calculate $L_{ce}$ as Eq. (\ref{equation3})
			\STATE Update the network's parameters:
			$ \Theta=\Theta+\frac{\partial{L_{ce}}}{\partial{\Theta}}$
			\ENDWHILE
			\STATE Calculate soft label $P_i$ as Eq (\ref{equation5})\\
			\STATE Send soft label $P_i$ to the information aggregation unit\\
			\REQUIRE {INFORMATION FUSION:}
			\STATE Aggregate knowledge: $\overrightarrow{P}=\frac{1}{\Gamma}\sum_{i=0}^{\Gamma}{P_i}$
			\REQUIRE {DISTILLATION TRAINING:}
			\WHILE {t in range(d)}
			\STATE Calculate $L_{ce}$ as Eq. (\ref{equation3})
			\STATE Calculate $L_{kl}$ as Eq. (\ref{equation7})
			\STATE The final loss: $L_d=L_{ce}+L_{kl}$
			\STATE Update the network's parameters:
			$ \Theta=\Theta+\frac{\partial{L_d}}{\partial{\Theta}}$
			\ENDWHILE
			\STATE q=q+1\\
			\STATE Update learning rate
			\ENDFOR\\
			
		\end{algorithmic}
	\end{algorithm}

	\section{Experiment}
	
	Two widely-used datasets are applied to verify the efficacy of our distillation framework, (1) the FSDKaggle2018 audio tagging dataset and (2) the 2018 TUT Urban acoustic scene classification dataset.
	
	\subsection{Dataset}
	\textbf{FSDKaggle2018 dataset.}
	The FSDKaggle2018 dataset \cite{fonseca2018general} was adopted for the general-purpose audio tagging task in 2018 DCASE, which aims to explore efficient models for general-purpose audio tagging problem. The samples in this dataset were annotated by Freesound \cite{fonseca2017freesound} with 41 labels (from Google’s AudioSet Ontology). The data format is unified to mono audio files of PCM 16 bit, frequency 44.1 KHz. There are about 9.5k samples in the training set, and 1.6K samples manually-verified annotated in the test set. Among the training set, samples are unequally distributed, which consists of about 3.7K manually-verified samples and about 5.8K non-verified samples whose quality estimated to be around 65-70\%. The sample clips range from 92 to 300 for different classes in the training set, while the duration of audio files differs from 300ms to 30s.

	\textbf{TUT Urban acoustic scenes 2018 dataset.}
	The TUT Urban Acoustic Scenes 2018 development dataset \cite{mesaros2018multi} (the dataset for DCASE 2018 task 1) for the acoustic scene classification task was used in our experiments, which tries to classify the characterizes of the environment where a recording from. Every sample in the training set is 10 seconds segments, and all of them are divided into 10 categories (acoustic scenes). Each acoustic scene contains 864 segments, in total 8640 segment. In the dataset, 6122 segments are used for training and 2518 segments are used for testing.

	\subsection{Networks}
	
	Convolutional neural networks (CNN) have shown their superiority in the audio classification tasks. However, few researchers have explored the application of the knowledge distillation method to the acoustic CNN model. In this paper, two representative CNN networks, VGGNet \cite{vgg} and ResNet \cite{he2016deep}, were used for experimental verification.
	
	\textbf{VGGNet} replaces the large convolution kernel (such as 7$\times$7 in AlexNet \cite{krizhevsky2012imagenet}) with a 3$\times$3 convolution kernel and improves its performance by deepening the network architecture. Three 3$\times$3 convolutional layers connected in series have the same effect with one 7$\times$7 convolutional layer, that is, the three 3$\times$3 convolutional layers have a receptive field size equivalent to one 7$\times$7 convolutional layer. However, the former has only about half of the latter's parameters and reducing linear operations which enhances the learning ability of models.
	
	\textbf{ResNet} uses the residual connection to solve the problems of information loss and vanishing gradients while training deep networks. The use of shortcut connections in ResNet directly bypasses the input information to the output, which protecting the integrity of the information and simplifying the learning objectives.
	
	In this paper, we use the classic 19-layer VGGNet network VGGNet19 and the 101-layer ResNet network ResNet101 in the distillation framework \cite{xu2018meta}.

	\subsection{Experiment setting}
	
	We use Pytorch to implement the network architecture and the librosa\footnote{https://github.com/librosa/librosa} toolkit package for data processing. In addition, the gRPC\footnote{https://github.com/grpc/grpc} framework is used for information transfer between networks. All the experiments are conducted on NVIDIA GeForce GTX 1080Ti GPU.
	For the experiment setting, SGD algorithm was adopted with the learning rate initialized as 0.001. The learning rate decays according to Pytorch CosineAnnealingLR function. The mini batch size set as 64 and the number of train epoch is 150. The mixup-data augmentation method \cite{zhang2018mixup} was adopted in all experiments to avoid overfitting. Setting the single branch training epoch $s=1$ and distillation training epoch $d=1$. In all experiments the models loading the parameters of pre-trained on ImageNet dataset.
	
	It is worthwhile to notice that many different audio representations can be deployed for the experiments. In this paper, log Mel, CQT and MFCC are used, while it is trial to extent our approach to other representations. To produce logMel and MFCC, we follow the setting of \cite{kelearticle}, the mel filter banks as 64, frameshift as 10 ms and frame width as 80 ms. Thus there will be 150 frames in an audio clip. The delta and delta-delta features are calculated using a window size of 9. To determine the relationship between features at different scales, logMel features of different resolution were used. In this paper, the logMel feature determined as "logMel128" where the number of mel filter banks is 128, and for the logMel feature whose number of mel filter banks is 64 determined as "logMel64". And in the tables without ambiguity, "logMel" defaults to logMel feature whose number of mel filter banks is 64.

	\section{Results and analysis}
	
	In this part, we present our experimental results of different configurations. In our experiments, the mean average precision (mAP) and accuracy were applied as the main evaluation criterion. All results reported are the audio level scores.
	
	%\subsubsection{Distillation using cross resolution representations.}
	
	\subsection{Knowledge distillation using cross multi-resolution representations}
	Table \ref{tab1} compares the results of networks trained with our distillation framework (masked with *) using cross resolution representations as inputs and the results of networks training independently.
	We can observe that from the table: (1) The performance of network distilled using cross resolution representations are better compared to the independently trained networks. (2) The networks with inputs of logMel128 perform better than the networks with inputs of logMel64, which indicates that logMel128 maintains more information about audio signals than logMel64. The reason may be that more mel filter banks are beneficial to preserve the detailed information. (3) Although the complementary information has been used to improve the performance of each branch in the distillation framework, the ensembled network are better than the single branch network in the framework, which demonstrates that there is still complementary information works in the ensemble network.
	The two groups (network as ResNet and VGGNet) of experiments have the same trend. In addition, although the classification results based on the VGGNet network are not as good as the experimental group based on the ResNet, our method achieves a bigger performance improvement on the experiments based on VGGNet.
	Knowledge distillation using cross resolution representations can be beneficial compared to the network training with the conventional method, indicating that different resolution representations enable networks to learn useful features to reach sufficient agreement.
	\begin{table}[!ht]
		\caption{The results (\%) of distillation using cross resolution representations with network architecture as ResNet on FSDKaggle2018 dataset. In the table, * indicates the branch network in our knowledge distillation framework, while network not distilled does not indicate with *. The Ensemble* is the results of ensemble networks of the branches in the framework. %\textbf{bf} represents the indicator was improved.
			The same in the following tables. \label{tab1}}
		\centering
		%% \tablesize{} %% You can specify the fontsize here, e.g., \tablesize{\footnotesize}. If commented out \small will be used.
		\begin{tabular}{c|c|c|c}
			\toprule%& \textbf{FLOPs}	& \textbf{Memory}
			\textbf{Network}	&\textbf{Input}	&\textbf{Accuracy}& \textbf{mAP@3}  \\
			\midrule
			\multirow{5}*{\textbf{ResNet}}	& logMel64 & 88.01&91.06\\
			~	&logMel128 &88.42&91.75 \\
			~	&logMel64* &89.43 &92.65 \\
			~	&logMel128* & 90.12& 93.16\\
			~	&Ensemble* &\textbf{91.02}&\textbf{93.84} \\
			\midrule
			\multirow{5}*{\textbf{VGGNet}}	& logMel64 & 82.4& 87.58\\
			~	&logMel128 & 83.12 & 88.45 \\
			~	&logMel64* & 89.75  & 92.66  \\
			~	&logMel128* & 89.81 & 92.98\\
			~	&Ensemble* &\textbf{90.5}&\textbf{93.48} \\
			\bottomrule
		\end{tabular}
	\end{table}

	\subsection{Knowledge distillation using multiple representations}
	To verify the effectiveness of multiple representations distillation, we use ResNet as the two branches' network architecture, while MFCC and LogMel are used as inputs, respectively. Table \ref{tab2} reports the accuracy and mAP@3 on the FSDKaggle2018 dataset of ResNet trained with independent training method and the results of distillation using multiple representations. And Table \ref{tab2-2} reports the results of distillation on multiple representations on the THU Urban Acoustic Scenes 2018 dataset.
	From the table, we can conclude that our distillation framework can leverage the complementary information in different representations to enhance the performance. It also can be seen that the branch with inputs of MFCC is better improved than the branch with inputs of logMel, more useful knowledge flow from the logMel branch to the MFCC branch during the distillation process. This is noteworthy that the basic performance of the logMel branch is better than the MFCC branch, that means that the branch of low-performance can get more performance gain from the high-performance branch.
	Another noteworthy phenomenon is that the accuracy and the mAP@3 of the logMel branch improved more after knowledge distillation with CQT than distillation with MFCC. This is in line with our expectations, because the conversion process from original audio to MFCC and logMel is similar, resulting in fewer feature differences between them. The CQT and logMel have more complementary information, which leads to a better effect of knowledge distillation.
	
	\begin{table}[!ht]
		\caption{The results (\%) of distillation using multiple representations with ResNet as network architecture on FSDKaggle2018 dataset. \label{tab2}}
		\centering
		%% \tablesize{} %% You can specify the fontsize here, e.g., \tablesize{\footnotesize}. If commented out \small will be used.
		\begin{tabular}{c|c|c|c}
			\toprule%& \textbf{FLOPs}	& \textbf{Memory}
			\textbf{Model}	&\textbf{Input}	&\textbf{Accuracy}& \textbf{mAP@3}  \\
			\midrule
			\multirow{5}*{\textbf{ResNet}}	&logMel & 88.01&91.06\\
			~	&MFCC &84.18&88.78\\
			~	&logMel* &88.31 &91.95 \\
			~	&MFCC* & 87.19 &90.86\\
			~	&Ensemble* &\textbf{90.44}&\textbf{93.28} \\

			\midrule
			\multirow{5}*{\textbf{ResNet}}	&logMel & 88.01&91.06\\
			~	&CQT &85.81&89.68\\
			~	&logMel* &89.63 &92.72 \\
			~	&CQT* & 87.69 &91.29\\
			~	&Ensemble* &\textbf{91.06}&\textbf{93.76} \\
			\bottomrule
		\end{tabular}
	\end{table}

	\begin{table}[!ht]
		\caption{The results (\%) of distillation using multiple representations with ResNet on TUT Urban acoustic scenes 2018 dataset. \label{tab2-2}}
		\centering
		%% \tablesize{} %% You can specify the fontsize here, e.g., \tablesize{\footnotesize}. If commented out \small will be used.
		\begin{tabular}{c|c|c|c}
			\toprule%& \textbf{FLOPs}	& \textbf{Memory}
			\textbf{Model}	&\textbf{Input}	&\textbf{Accuracy}& \textbf{mAP@3}  \\
			\midrule
			\multirow{5}*{\textbf{VGGNet}}	&logMel & 65.29& 76.88\\
			~	&MFCC &63.46 & 75.11 \\
			~	&logMel* & 66.12   &77.44  \\
			~	&MFCC* & 65.8 & 76.22\\
			~	&Ensemble* &\textbf{66.76}&\textbf{77.7} \\
			\midrule
			\multirow{5}*{\textbf{ResNet}}	&logMel & 70.33& 80.67\\
			~	&MFCC &67.71 & 78.32 \\
			~	&logMel* &  72.43  &81.65  \\
			~	&MFCC* & 68.59 &79.82\\
			~	&Ensemble* &\textbf{72.86}&\textbf{82.25} \\

			\bottomrule
		\end{tabular}
	\end{table}

	\subsection{Knowledge distillation using different network architectures}
	
	As the different inputs could produce complementary information, the network architectures may also lead to differences in knowledge. Table \ref{tab3} and Table \ref{tab3-2} compare the results in the distillation framework which distilled with ResNet and VGGNet network and the results of independent training method on TUT Urban acoustic scenes 2018 dataset and FSDKaggle2018 dataset. As expected, the knowledge distillation framework provided sufficient promotion compared to the independently training. Different network structures are also sources of complementary knowledge.
	Although our framework is useful on both the FSDKaggle2018 dataset (for general-purpose audio tagging task) and the TUT Urban acoustic scenes 2018 dataset (for acoustic scenes classification task), it is apparent that the promotion is greater on the general-purpose audio tagging task. There are a large number of non-verified samples on the FSDKaggle2018 dataset, and in distillation the soft labels re-marking the erroneous data can greatly reduce the impact of the error label. The process of relabeling the target by the knowledge distillation method can obtain similar benefits to semi-supervised learning and reduce the false induction of confusing labels.

	\begin{table}[!ht]\setlength{\tabcolsep}{2pt}
		\caption{The results (\%) of distillation using different network architectures with inputs of logMel on TUT Urban acoustic scenes 2018 dataset. \label{tab3}}
		\centering
		%% \tablesize{} %% You can specify the fontsize here, e.g., \tablesize{\footnotesize}. If commented out \small will be used.
		\begin{tabular}{c|c|c|c}
			\toprule%& \textbf{FLOPs}	& \textbf{Memory}
			{\textbf{Input}}&{\textbf{Network}} &\textbf{Accuracy}& \textbf{mAP@3}\\
			\midrule
			\multirow{5}*{logMel}&VGGNet & 65.29&	76.88\\
			~&ResNet &70.33&	80.67\\
			~&VGGNet*  &66.6&	77.59\\
			~&ResNet*& 70.49& 80.51\\
			~&Ensemble*& \textbf{70.84}&\textbf{80.85}\\
			\bottomrule
		\end{tabular}
	\end{table}

	\begin{table}[!ht]%\setlength{\tabcolsep}{5pt}
		\caption{The results (\%) of distillation using different network architectures with inputs of CQT on FSDKaggle2018 dataset. \label{tab3-2}}
		\centering
		%% \tablesize{} %% You can specify the fontsize here, e.g., \tablesize{\footnotesize}. If commented out \small will be used.
		\begin{tabular}{c|c|c|c}
			\toprule%& \textbf{FLOPs}	& \textbf{Memory}
			{\textbf{Input}} &{\textbf{Network}} &\textbf{Accuracy}& \textbf{mAP@3} \\
			\midrule
			\multirow{5}*{logMel}&VGGNet & 82.4&	87.58\\
			~&ResNet &88.01&91.06 \\
			~&VGGNet* &86.43&90.58 \\
			~&ResNet* & 89.18&92.61\\
			~&Ensemble* & \textbf{89.88}&\textbf{93.05}\\
			\midrule
			\multirow{5}*{\textbf{CQT}}&VGGNet & 75.13&	82.55\\
			~	&ResNet &85.81&89.68\\
			~	&VGGNet* &86.2&	90.42\\
			~	&ResNet* & 87.94&91.31\\
			~	&Ensemble* & \textbf{89.93}&\textbf{93.18}\\
			\bottomrule
		\end{tabular}
	\end{table}

	\begin{table*}[!ht]\setlength{\tabcolsep}{5pt}
		\caption{The results (\%) of distillation using multiple representations and different network architectures on FSDKaggle2018 dataset. I is an abbreviation for input, and N is an abbreviation for network. \label{tab5}}
		\centering
		%% \tablesize{} %% You can specify the fontsize here, e.g., \tablesize{\footnotesize}. If commented out \small will be used.
		\begin{tabular}{ccccc|ccccc|cc}
			\toprule%& \textbf{FLOPs}	& \textbf{Memory}
			\multicolumn{5}{c|}{Branch1}&\multicolumn{5}{c}{Branch2}&\multicolumn{2}{|c}{Ensemble}\\
			
			\textbf{I1,N1}	&\textbf{ACC}&\textbf{mAP@3}&\textbf{ACC*}&\textbf{mAP@3*}& 	\textbf{I2,N2} &\textbf{ACC}&\textbf{mAP@3} &\textbf{ACC*}&\textbf{mAP@3*}&\textbf{ACC*}&\textbf{mAP@3*} \\
			\midrule
			logMel,ResNet &88.01&91.06&90.18 &93.26 &MFCC,VGGNet&81.54&86.88&84.31&88.96&\textbf{90.69}&\textbf{93.27}\\
			MFCC,ResNet &84.18&	88.78&87.5 &91.13 &logMel,VGGNet&82.4&87.58&89.5&92.78&\textbf{89.93}&\textbf{93.18}\\
			logMel,ResNet &88.01&91.06&90.6 &93.16 &CQT,VGGNet&75.12&82.55&86.5&90.73&\textbf{91.31}&\textbf{94.11}\\
			\bottomrule
		\end{tabular}
	\end{table*}
	
	\begin{table*}[!ht]\setlength{\tabcolsep}{5pt}
		\caption{The results (\%) of distillation using multiple representations and different network architectures on TUT Urban acoustic scenes 2018 dataset. I is an abbreviation for input, and N is an abbreviation for network. \label{tab6}}
		\centering
		%% \tablesize{} %% You can specify the fontsize here, e.g., \tablesize{\footnotesize}. If commented out \small will be used.
		\begin{tabular}{ccccc|ccccc|cc}
			\toprule%& \textbf{FLOPs}	& \textbf{Memory}
			\multicolumn{5}{c|}{Branch1}&\multicolumn{5}{c}{Branch2}&\multicolumn{2}{|c}{Ensemble}\\
			
			\textbf{I1,N1}	&\textbf{ACC}&\textbf{mAP@3}&\textbf{ACC*}&\textbf{mAP@3*}& 	\textbf{I2,N2}&\textbf{ACC}&\textbf{mAP@3} &\textbf{ACC*}&\textbf{mAP@3*} &\textbf{ACC*}&\textbf{mAP@3*} \\
			\midrule
			logMel,ResNet &70.33	&80.67&\textbf{72.48} &\textbf{81.65} &MFCC,VGGNet&63.46&	75.11&64.29&75.24&71.17&80.98\\
			MFCC,ResNet &67.71&	78.32& 67.91 &77.87 &logMel,VGGNet&65.29& 76.88&68.19&78.38& \textbf{70.29}&\textbf{80.14}\\
			\bottomrule
		\end{tabular}
	\end{table*}

	\subsection{Knowledge distillation using multiple representations and different network architectures}
	Further experiments explored the impact of multi-representations and different network architectures on knowledge distillation. Table \ref{tab5} and Table \ref{tab6} report the results where two branches use completely different inputs and network architectures on the general-purpose audio tagging task and acoustic scenes classification task, respectively.
	Theoretically, the more different factors are set in the branches, the more complementary information produced. From Table \ref{tab5}, we can find that the branches distilled using multi-representations and different network architectures are much better than trained independent. And the improvements (both accuracy and mAP@3) of distilled branches are much bigger than distillation just using multiple representations or different architectures. This indicates that the difference between the branches no matter input or network architecture directly determines effect of distillation. In Table \ref{tab6}, similar trends as Table \ref{tab5} can be observed. These conclusions also hold on the TUT Urban acoustic scenes 2018 dataset. In addition, in the first set of experiments in TUT Urban acoustic scenes 2018 dataset, the ensemble collapse phenomenon occurred for the reason of the large performance gap of the branch networks, which indicates that our knowledge distillation method is more stable than the ensemble method.

	\begin{figure}[!ht]
		\centering
		\includegraphics[scale=0.6]{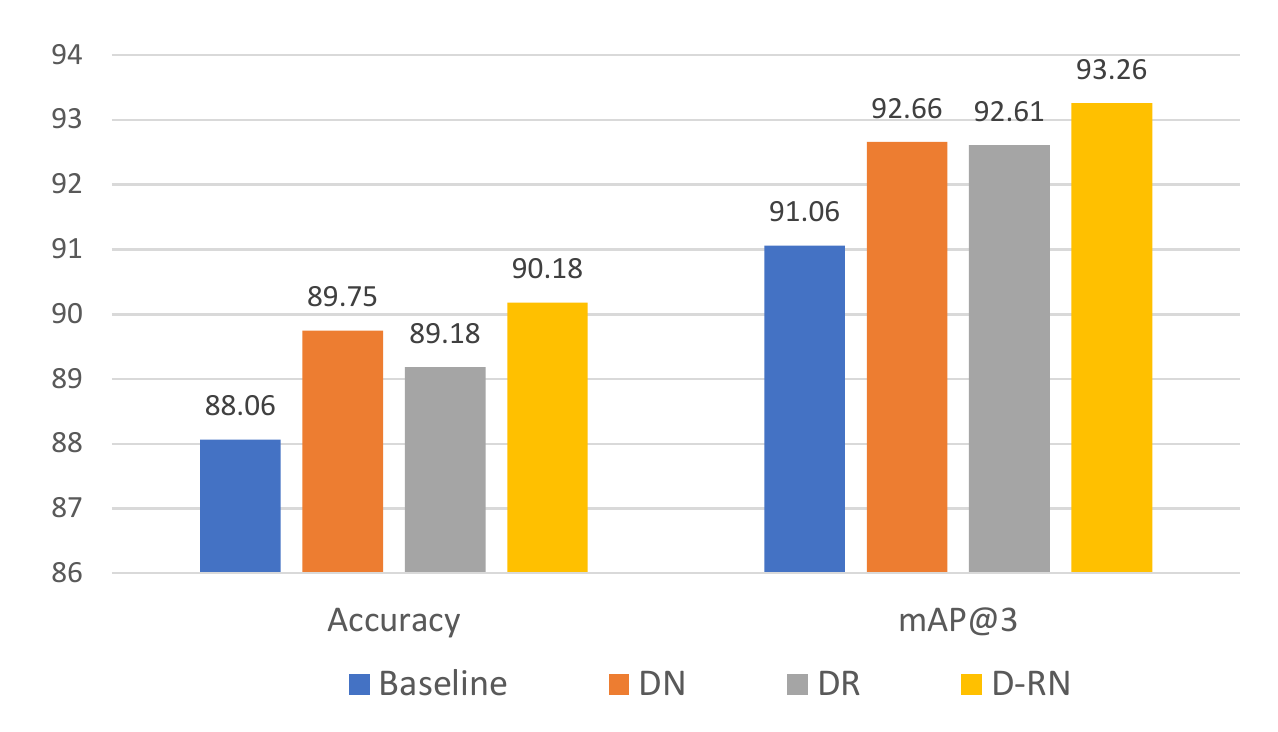}
		\caption{The results (\%) comparison of different mode distillation on FSDKaggle2018 dataset, we only report the results of the branch with logMel as input and ResNet as the network in distillation framework. DN is an abbreviation for distillation using different network architectures, DR is an abbreviation for distillation using different representations, and D-RN stands for distillation using different representations and different network architectures.}
		\label{fig3}
	\end{figure}
	
	\subsection{The results comparison of distillation in different settings}
	Figure \ref{fig3} compare the results in different setting of our distillation framework. We can find that the networks with our knowledge distillation still perform better that the results of baseline (independently trained network). And it is obvious that the results of distillation using multiple representations and different network architectures are better than only distillation using multiple representations or distillation using different network architectures. From the all above results we conclude that (1) The knowledge distillation method always improves the performance of the branch network, and the ensemble method works for the distilled branches. (2) Both multiple representations distillation and different architectures provided supplementary information needed for distillation. (3) The biggest improvement comes from the distillation simultaneous different representations and different architectures, where branch network gets the best performance. This suggests that complementary knowledge from different sources can be superimposed.
	
	\begin{figure}[!ht]
		\centering
		\includegraphics[width=8cm,height=5cm]{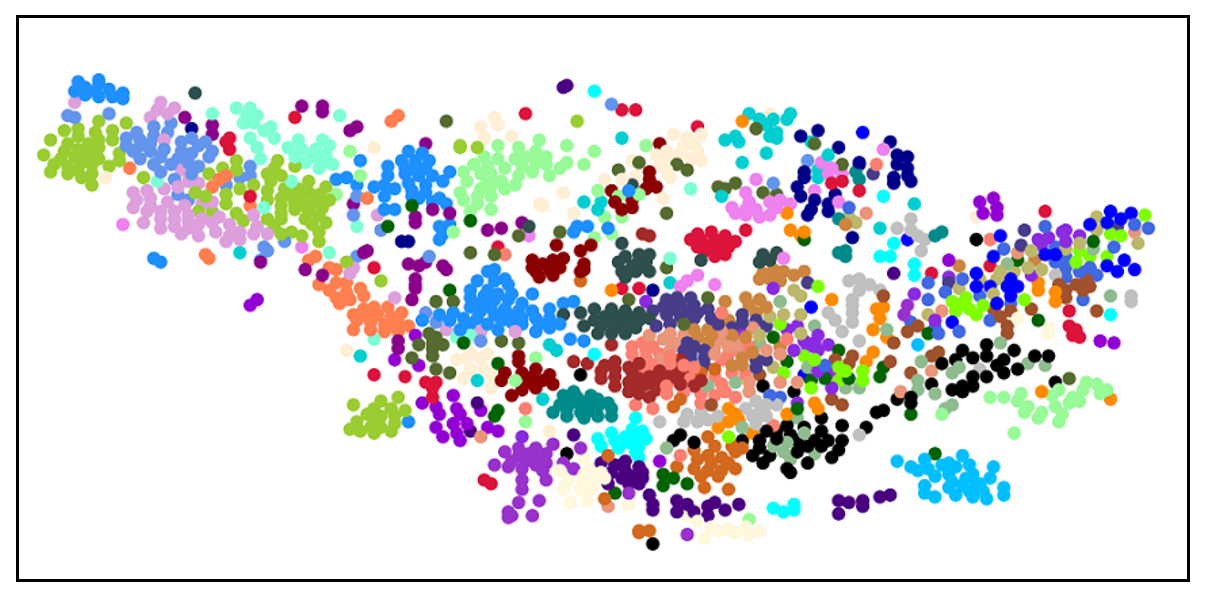}
		\caption{The T-SNE visualization figure without using our method.}
		\label{fig4}
	\end{figure}
	\begin{figure}[!ht]
		\centering
		\includegraphics[width=8cm,height=5cm]{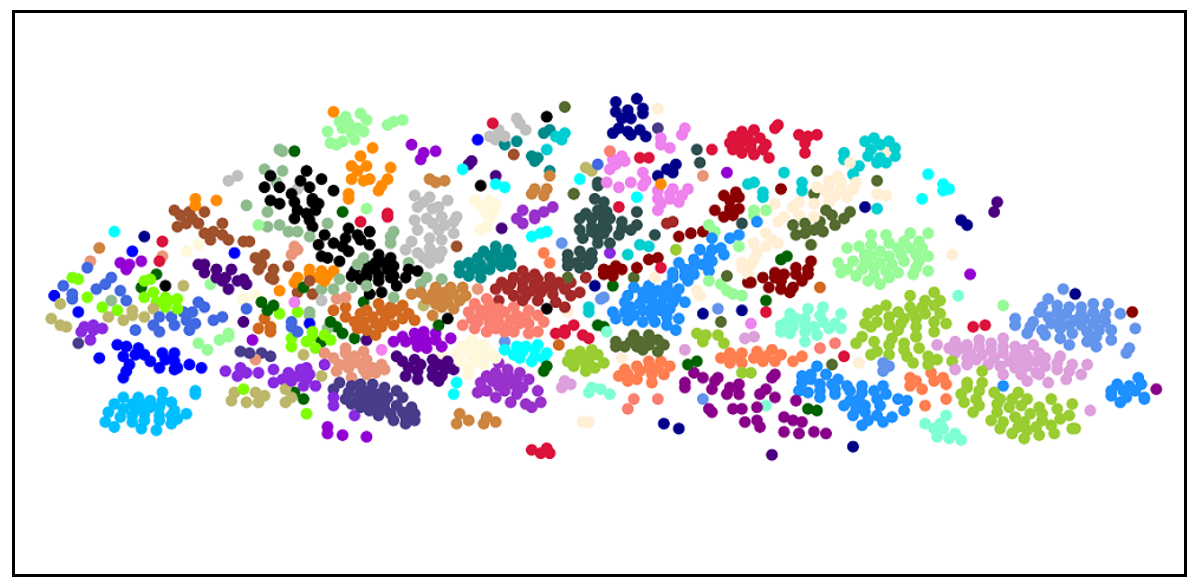}
		\caption{The T-SNE visualization figure of using our method.}
		\label{fig5}
	\end{figure}
	
	\subsection{The T-SNE visualization analysis}

	To show the effectiveness of our method clearly, we using the t-distributed Stochastic Neighbor (T-SNE) embedding visualization method. The T-SNE method is an efficient manifold learning method that can compress high-dimensional data to a low-dimensional structure. Figures \ref{fig4} shows the T-SNE visualization figure of a ResNet network (CQT as input) trained independently and Figures \ref{fig5} was the results with our multi-representation distillation method. We used the 1600 manually verified samples on the validation set of the THU 2018 dataset, and got the logits layer outputs of the network, then using the T-SNE method to compress logits outputs into two-dimensional space to visualize the results. It can be found that without knowledge distillation, the points of samples are more scattered and there are many crossovers between samples of different categories, which makes it more difficult to distinguish the category boundaries. In contrast, in the T-SNE visualization figure of the model trained by our knowledge distillation method, the samples are compact and the sample category confusion is reduced. The figures show that our method can effectively enhance the classification performance of audio classification models.

	\section{Conclusions}
	In this paper, we propose a novel collaborative learning framework for the audio classification task. It takes multiple representations as input and trains a classifier separately on each representation. A collaborative distillation framework is employed to share knowledge across different models. Extensive experiments demonstrate that the proposed approach can improve the classification performance and achieve competitive results on both acoustic scenes classification task and general audio tagging task (experiments were conducted on the FSD-Kaggle2018 dataset and the TUT Urban acoustic scenes classification 2018 dataset). Moreover, it is worthwhile to notice that leveraging this approach is capable of promoting the performance of the model, without increasing the computational complexity in the inference phase. A direction of future work is using the multi-representation distillation method to improve the performance on the tasks of the sound event detection. Moreover, we would like to explore of knowledge distillation for multi-modal datasets.

	\section*{Acknowledgment}

	The authors would like to thank the National Key R\&D Program of China (2016YFB1000101).

	% or
	

	\appendices

	\ifCLASSOPTIONcaptionsoff
	\newpage
	\fi

	
	
	%\bibliographystyle{IEEEtran}
	
	%\bibliographystyle{ieeetr}
	\bibliographystyle{IEEEtran}
	\bibliography{IEEEabrv,reference}

\begin{thebibliography}{10}
\providecommand{\url}[1]{#1}
\csname url@samestyle\endcsname
\providecommand{\newblock}{\relax}
\providecommand{\bibinfo}[2]{#2}
\providecommand{\BIBentrySTDinterwordspacing}{\spaceskip=0pt\relax}
\providecommand{\BIBentryALTinterwordstretchfactor}{4}
\providecommand{\BIBentryALTinterwordspacing}{\spaceskip=\fontdimen2\font plus
\BIBentryALTinterwordstretchfactor\fontdimen3\font minus
  \fontdimen4\font\relax}
\providecommand{\BIBforeignlanguage}[2]{{%
\expandafter\ifx\csname l@#1\endcsname\relax
\typeout{** WARNING: IEEEtran.bst: No hyphenation pattern has been}%
\typeout{** loaded for the language `#1'. Using the pattern for}%
\typeout{** the default language instead.}%
\else
\language=\csname l@#1\endcsname
\fi
#2}}
\providecommand{\BIBdecl}{\relax}
\BIBdecl

\bibitem{virtanen2018computational}
T.~Virtanen, M.~D. Plumbley, and D.~Ellis, \emph{Computational analysis of
  sound scenes and events}.\hskip 1em plus 0.5em minus 0.4em\relax
  Heidelberg:Springer, 2018.

\bibitem{clavel2005events}
C.~Clavel, T.~Ehrette, and G.~Richard, ``Events detection for an audio-based
  surveillance system,'' in \emph{IEEE International Conference on Multimedia
  and Expo}.\hskip 1em plus 0.5em minus 0.4em\relax IEEE, 2005, pp. 1306--1309.

\bibitem{peng2009healthcare}
Y.-T. Peng, C.-Y. Lin, M.-T. Sun, and K.-C. Tsai, ``Healthcare audio event
  classification using hidden markov models and hierarchical hidden markov
  models,'' in \emph{2009 IEEE International Conference on Multimedia and
  Expo}.\hskip 1em plus 0.5em minus 0.4em\relax IEEE, 2009, pp. 1218--1221.

\bibitem{ma2003context}
L.~Ma, D.~Smith, and B.~Milner, ``Context awareness using environmental noise
  classification,'' in \emph{European Conference on Speech Communication and
  Technology}, 2003, pp. 2237--2240.

\bibitem{xu2018large}
Y.~Xu, Q.~Kong, W.~Wang, and M.~D. Plumbley, ``Large-scale weakly supervised
  audio classification using gated convolutional neural network,'' in
  \emph{International Conference on Acoustics, Speech and Signal
  Processing}.\hskip 1em plus 0.5em minus 0.4em\relax IEEE, 2018, pp. 121--125.

\bibitem{dhanalakshmi2009classification}
P.~Dhanalakshmi, S.~Palanivel, and V.~Ramalingam, ``Classification of audio
  signals using svm and rbfnn,'' \emph{Expert systems with applications},
  vol.~36, no.~3, pp. 6069--6075, 2009.

\bibitem{lee2006automatic}
K.~Lee and M.~Slaney, ``Automatic chord recognition from audio using a hmm with
  supervised learning.'' in \emph{ISMIR}, 2006, pp. 133--137.

\bibitem{dhanalakshmi2011classification}
P.~Dhanalakshmi, S.~Palanivel, and V.~Ramalingam, ``Classification of audio
  signals using aann and gmm,'' \emph{Applied Soft Computing}, vol.~11, no.~1,
  pp. 716--723, 2011.

\bibitem{sainath2015learning}
T.~N. Sainath, R.~J. Weiss, A.~Senior, K.~W. Wilson, and O.~Vinyals, ``Learning
  the speech front-end with raw waveform cldnns,'' in \emph{Sixteenth Annual
  Conference of the International Speech Communication Association}, 2015.

\bibitem{choi2017convolutional}
K.~Choi, G.~Fazekas, M.~Sandler, and K.~Cho, ``Convolutional recurrent neural
  networks for music classification,'' in \emph{2017 IEEE International
  Conference on Acoustics, Speech and Signal Processing (ICASSP)}.\hskip 1em
  plus 0.5em minus 0.4em\relax IEEE, 2017, pp. 2392--2396.

\bibitem{7100934}
D.~{Stowell}, D.~{Giannoulis}, E.~{Benetos}, M.~{Lagrange}, and M.~D.
  {Plumbley}, ``Detection and classification of acoustic scenes and events,''
  \emph{IEEE Transactions on Multimedia}, vol.~17, no.~10, pp. 1733--1746, Oct
  2015.

\bibitem{mesaros2018detection}
A.~Mesaros, T.~Heittola, E.~Benetos, P.~Foster, M.~Lagrange, T.~Virtanen, and
  M.~D. Plumbley, ``Detection and classification of acoustic scenes and events:
  Outcome of the dcase 2016 challenge,'' \emph{IEEE/ACM Transactions on Audio,
  Speech and Language Processing}, vol.~26, no.~2, pp. 379--393, 2018.

\bibitem{lee2018samplecnn}
J.~Lee, J.~Park, K.~L. Kim, and J.~Nam, ``Samplecnn: End-to-end deep
  convolutional neural networks using very small filters for music
  classification,'' \emph{Applied Sciences}, vol.~8, no.~1, p. 150, 2018.

\bibitem{lee2017sample-level}
------, ``Sample-level deep convolutional neural networks for music
  auto-tagging using raw waveforms.'' \emph{arXiv: Sound}, 2017.

\bibitem{piczak2015environmental}
K.~J. Piczak, ``Environmental sound classification with convolutional neural
  networks,'' in \emph{2015 IEEE 25th International Workshop on Machine
  Learning for Signal Processing (MLSP)}.\hskip 1em plus 0.5em minus
  0.4em\relax IEEE, 2015, pp. 1--6.

\bibitem{8085174}
S.~{Zhang}, S.~{Zhang}, T.~{Huang}, and W.~{Gao}, ``Speech emotion recognition
  using deep convolutional neural network and discriminant temporal pyramid
  matching,'' \emph{IEEE Transactions on Multimedia}, vol.~20, no.~6, pp.
  1576--1590, June 2018.

\bibitem{sercu2016dense}
T.~Sercu and V.~Goel, ``Dense prediction on sequences with time-dilated
  convolutions for speech recognition,'' \emph{arXiv preprint
  arXiv:1611.09288}, 2016.

\bibitem{sainath2015convolutional}
T.~N. Sainath and C.~Parada, ``Convolutional neural networks for
  small-footprint keyword spotting,'' in \emph{Sixteenth Annual Conference of
  the International Speech Communication Association}, 2015.

\bibitem{hoshen2015speech}
Y.~Hoshen, R.~J. Weiss, and K.~W. Wilson, ``Speech acoustic modeling from raw
  multichannel waveforms,'' in \emph{2015 IEEE International Conference on
  Acoustics, Speech and Signal Processing (ICASSP)}.\hskip 1em plus 0.5em minus
  0.4em\relax IEEE, 2015, pp. 4624--4628.

\bibitem{zhang2019large}
L.~Zhang, D.~Wang, C.~Bao, Y.~Wang, and K.~Xu, ``Large-scale whale-call
  classification by transfer learning on multi-scale waveforms and
  time-frequency features,'' \emph{Applied Sciences}, vol.~9, no.~5, p. 1020,
  2019.

\bibitem{wang2019acoustic}
D.~Wang, L.~Zhang, K.~Xu, and Y.~Wang, ``Acoustic scene classification based on
  dense convolutional networks incorporating multi-channel features,'' in
  \emph{Journal of Physics: Conference Series}, vol. 1169, no.~1.\hskip 1em
  plus 0.5em minus 0.4em\relax IOP Publishing, 2019, p. 012037.

\bibitem{ramalingam2006gaussian}
A.~Ramalingam and S.~Krishnan, ``Gaussian mixture modeling of short-time
  fourier transform features for audio fingerprinting,'' \emph{IEEE
  Transactions on Information Forensics and Security}, vol.~1, no.~4, pp.
  457--463, 2006.

\bibitem{yin2018learning}
Y.~Yin, R.~R. Shah, and R.~Zimmermann, ``Learning and fusing multimodal deep
  features for acoustic scene categorization,'' in \emph{ACM Multimedia
  Conference on Multimedia Conference}.\hskip 1em plus 0.5em minus 0.4em\relax
  ACM, 2018, pp. 1892--1900.

\bibitem{xu2019general}
K.~Xu, B.~Zhu, Q.~Kong, H.~Mi, B.~Ding, D.~Wang, and H.~Wang, ``General audio
  tagging with ensembling convolutional neural networks and statistical
  features,'' \emph{The Journal of the Acoustical Society of America}, vol.
  145, no.~6, pp. EL521--EL527, 2019.

\bibitem{xu2018meta}
K.~Xu, B.~Zhu, D.~Wang, Y.~Peng, H.~Wang, L.~Zhang, and B.~Li, ``Meta learning
  based audio tagging,'' in \emph{Proceedings of the Workshop on Detection and
  Classification of Acoustic Scenes and Events (DCASE 2018), Surrey, UK}, 2018,
  pp. 19--20.

\bibitem{8059841}
C.~{Zhang}, J.~{Cheng}, and Q.~{Tian}, ``Multiview label sharing for visual
  representations and classifications,'' \emph{IEEE Transactions on
  Multimedia}, vol.~20, no.~4, pp. 903--913, April 2018.

\bibitem{dietterich2000ensemble}
T.~G. Dietterich, ``Ensemble methods in machine learning,'' in
  \emph{International workshop on multiple classifier systems}.\hskip 1em plus
  0.5em minus 0.4em\relax Springer, 2000, pp. 1--15.

\bibitem{rokach2010ensemble}
L.~Rokach, ``Ensemble-based classifiers,'' \emph{Artificial Intelligence
  Review}, vol.~33, no. 1-2, pp. 1--39, 2010.

\bibitem{wang2018weakly}
D.~Wang, L.~Zhang, C.~Bao, K.~Xu, B.~Zhu, and Q.~Kong, ``Weakly supervised crnn
  system for sound event detection with large-scale unlabeled in-domain data,''
  \emph{arXiv preprint arXiv:1811.00301}, 2018.

\bibitem{8490588}
C.~{Ma}, Y.~{Guo}, J.~{Yang}, and W.~{An}, ``Learning multi-view representation
  with lstm for 3-d shape recognition and retrieval,'' \emph{IEEE Transactions
  on Multimedia}, vol.~21, no.~5, pp. 1169--1182, May 2019.

\bibitem{hinton2015distilling}
G.~Hinton, O.~Vinyals, and J.~Dean, ``Distilling the knowledge in a neural
  network,'' \emph{Stat}, vol. 1050, p.~9, 2015.

\bibitem{zhang2018deep}
Y.~Zhang, T.~Xiang, T.~M. Hospedales, and H.~Lu, ``Deep mutual learning,'' in
  \emph{Proceedings of the IEEE Conference on Computer Vision and Pattern
  Recognition}, 2018, pp. 4320--4328.

\bibitem{sun2017ensemble}
S.~Sun, W.~Chen, J.~Bian, X.~Liu, and T.-Y. Liu, ``Ensemble-compression: A new
  method for parallel training of deep neural networks,'' in \emph{Joint
  European Conference on Machine Learning and Knowledge Discovery in
  Databases}.\hskip 1em plus 0.5em minus 0.4em\relax Springer, 2017, pp.
  187--202.

\bibitem{burges1998tutorial}
C.~J. Burges, ``A tutorial on support vector machines for pattern
  recognition,'' \emph{Data Mining and Knowledge Discovery}, vol.~2, no.~2, pp.
  121--167, 1998.

\bibitem{phan2016label}
H.~Phan, L.~Hertel, M.~Maass, P.~Koch, and A.~Mertins, ``Label tree embeddings
  for acoustic scene classification,'' in \emph{ACM international conference on
  Multimedia}.\hskip 1em plus 0.5em minus 0.4em\relax ACM, 2016, pp. 486--490.

\bibitem{dietterich2000experimental}
T.~G. Dietterich, ``An experimental comparison of three methods for
  constructing ensembles of decision trees: Bagging, boosting, and
  randomization,'' \emph{Machine Learning}, vol.~40, no.~2, pp. 139--157, 2000.

\bibitem{kulkarni2009audio}
A.~Kulkarni, ``Audio signal processing,'' Feb.~10 2009, uS Patent 7,490,044.

\bibitem{aytar2016soundnet}
Y.~Aytar, C.~Vondrick, and A.~Torralba, ``Soundnet: Learning sound
  representations from unlabeled video,'' in \emph{Advances in Neural
  Information Processing Systems}, 2016, pp. 892--900.

\bibitem{huang2017densely}
G.~Huang, Z.~Liu, L.~Van Der~Maaten, and K.~Q. Weinberger, ``Densely connected
  convolutional networks.'' in \emph{IEEE Conference on Computer Vision and
  Pattern Recognition}, vol.~1, 2017, p.~3.

\bibitem{he2016deep}
K.~He, X.~Zhang, S.~Ren, and J.~Sun, ``Deep residual learning for image
  recognition,'' in \emph{Proceedings of the IEEE Conference on Computer Vision
  and Pattern Recognition}, 2016, pp. 770--778.

\bibitem{zhu2018environmental}
B.~Zhu, K.~Xu, D.~Wang, L.~Zhang, B.~Li, and Y.~Peng, ``Environmental sound
  classification based on multi-temporal resolution convolutional neural
  network combining with multi-level features,'' in \emph{Pacific Rim
  Conference on Multimedia}.\hskip 1em plus 0.5em minus 0.4em\relax Springer,
  2018, pp. 528--537.

\bibitem{poggio1990regularization}
T.~Poggio and F.~Girosi, ``Regularization algorithms for learning that are
  equivalent to multilayer networks,'' \emph{Science}, vol. 247, no. 4945, pp.
  978--982, 1990.

\bibitem{xu2018mixup}
K.~Xu, D.~Feng, H.~Mi, B.~Zhu, D.~Wang, L.~Zhang, H.~Cai, and S.~Liu,
  ``Mixup-based acoustic scene classification using multi-channel convolutional
  neural network,'' in \emph{Pacific Rim Conference on Multimedia}.\hskip 1em
  plus 0.5em minus 0.4em\relax Springer, 2018, pp. 14--23.

\bibitem{wei2018sample}
S.~Wei, K.~Xu, D.~Wang, F.~Liao, H.~Wang, and Q.~Kong, ``Sample mixed-based
  data augmentation for domestic audio tagging,'' in \emph{DCASE 2018
  Workshop}, 2018.

\bibitem{feng2018sample}
D.~Feng, K.~Xu, H.~Mi, F.~Liao, and Y.~Zhou, ``Sample dropout for audio scene
  classification using multi-scale dense connected convolutional neural
  network,'' in \emph{Pacific Rim Knowledge Acquisition Workshop}.\hskip 1em
  plus 0.5em minus 0.4em\relax Springer, 2018, pp. 114--123.

\bibitem{aucouturier2007bag}
J.-J. Aucouturier, B.~Defreville, and F.~Pachet, ``The bag-of-frames approach
  to audio pattern recognition: A sufficient model for urban soundscapes but
  not for polyphonic music,'' \emph{The Journal of the Acoustical Society of
  America}, vol. 122, no.~2, pp. 881--891, 2007.

\bibitem{dehak2010front}
N.~Dehak, P.~J. Kenny, R.~Dehak, P.~Dumouchel, and P.~Ouellet, ``Front-end
  factor analysis for speaker verification,'' \emph{IEEE Transactions on Audio,
  Speech, and Language Processing}, vol.~19, no.~4, pp. 788--798, 2010.

\bibitem{7927482}
L.~{Jing}, B.~{Liu}, J.~{Choi}, A.~{Janin}, J.~{Bernd}, M.~W. {Mahoney}, and
  G.~{Friedland}, ``Dcar: A discriminative and compact audio representation for
  audio processing,'' \emph{IEEE Transactions on Multimedia}, vol.~19, no.~12,
  pp. 2637--2650, Dec 2017.

\bibitem{7592459}
J.~{Ren}, X.~{Jiang}, J.~{Yuan}, and N.~{Magnenat-Thalmann}, ``Sound-event
  classification using robust texture features for robot hearing,'' \emph{IEEE
  Transactions on Multimedia}, vol.~19, no.~3, pp. 447--458, March 2017.

\bibitem{kuncheva2003measures}
L.~I. Kuncheva and C.~J. Whitaker, ``Measures of diversity in classifier
  ensembles and their relationship with the ensemble accuracy,'' \emph{Machine
  Learning}, vol.~51, no.~2, pp. 181--207, 2003.

\bibitem{sollich1996learning}
P.~Sollich and A.~Krogh, ``Learning with ensembles: How overfitting can be
  useful,'' in \emph{Advances in Neural Information Processing Systems}, 1996,
  pp. 190--196.

\bibitem{bucilua2006model}
C.~Bucilua, R.~Caruana, and A.~Niculescu-Mizil, ``Model compression,'' in
  \emph{Proceedings of the 12th ACM SIGKDD International Conference on
  Knowledge Discovery and Data Mining}.\hskip 1em plus 0.5em minus 0.4em\relax
  ACM, 2006, pp. 535--541.

\bibitem{gao2019multistructure}
L.~Gao, X.~Lan, H.~Mi, D.~Feng, K.~Xu, and Y.~Peng, ``Multistructure-based
  collaborative online distillation,'' \emph{Entropy}, vol.~21, no.~4, p. 357,
  2019.

\bibitem{kim2018paraphrasing}
J.~Kim, S.~Park, and N.~Kwak, ``Paraphrasing complex network: Network
  compression via factor transfer,'' in \emph{Advances in Neural Information
  Processing Systems}, 2018, pp. 2760--2769.

\bibitem{anil2018large}
R.~Anil, G.~Pereyra, A.~Passos, R.~Ormandi, G.~E. Dahl, and G.~E. Hinton,
  ``Large scale distributed neural network training through online
  distillation,'' \emph{International Conference on Learning Representations},
  2018.

\bibitem{batra2017cooperative}
T.~Batra and D.~Parikh, ``Cooperative learning with visual attributes.''
  \emph{Computer Vision and Pattern Recognition}, 2017.

\bibitem{lan2018knowledge}
X.~Lan, X.~Zhu, and S.~Gong, ``Knowledge distillation by on-the-fly native
  ensemble,'' in \emph{Proceedings of the 32nd International Conference on
  Neural Information Processing Systems (NIPS)}.\hskip 1em plus 0.5em minus
  0.4em\relax Curran Associates Inc., 2018, pp. 7528--7538.

\bibitem{fonseca2018general}
E.~Fonseca, M.~Plakal, F.~Font, D.~P. Ellis, X.~Favory, J.~Pons, and X.~Serra,
  ``General-purpose tagging of freesound audio with audioset labels: Task
  description, dataset, and baseline,'' \emph{Proceedings of the Detection and
  Classification of Acoustic Scenes and Events Workshop}, p. 69–73, 2018.

\bibitem{fonseca2017freesound}
E.~Fonseca, J.~Pons~Puig, X.~Favory, F.~Font~Corbera, D.~Bogdanov, A.~Ferraro,
  S.~Oramas, A.~Porter, and X.~Serra, ``Freesound datasets: a platform for the
  creation of open audio datasets,'' in \emph{Hu X, Cunningham SJ, Turnbull D,
  Duan Z, editors. Proceedings of the 18th ISMIR Conference; 2017 oct 23-27;
  Suzhou, China.[Canada]: International Society for Music Information
  Retrieval; 2017. p. 486-93.}\hskip 1em plus 0.5em minus 0.4em\relax
  International Society for Music Information Retrieval, 2017.

\bibitem{mesaros2018multi}
T.~H. Annamaria~Mesaros and T.~Virtanen, ``A multi-device dataset for urban
  acoustic scene classification,'' \emph{Proceedings of the Detection and
  Classification of Acoustic Scenes and Events Workshop}, p. 9–13, 2018.

\bibitem{vgg}
K.~Simonyan and A.~Zisserman, ``Very deep convolutional networks for
  large-scale image recognition,'' in \emph{International Conference on
  Learning Representations}, May 2015.

\bibitem{krizhevsky2012imagenet}
A.~Krizhevsky, I.~Sutskever, and G.~E. Hinton, ``Imagenet classification with
  deep convolutional neural networks,'' in \emph{Advances in Neural Information
  Processing Systems}, 2012, pp. 1097--1105.

\bibitem{zhang2018mixup}
\BIBentryALTinterwordspacing
H.~Zhang, M.~Cisse, Y.~N. Dauphin, and D.~Lopez-Paz, ``mixup: Beyond empirical
  risk minimization,'' in \emph{International Conference on Learning
  Representations}, 2018. [Online]. Available:
  \url{https://openreview.net/forum?id=r1Ddp1-Rb}
\BIBentrySTDinterwordspacing

\bibitem{kelearticle}
K.~Xu, B.~Zhu, Q.~Kong, H.~Mi, B.~Ding, D.~Wang, and H.~Wang, ``General audio
  tagging with ensembling convolutional neural networks and statistical
  features,'' \emph{The Journal of the Acoustical Society of America}, vol.
  145, pp. 521--527, 06 2019.

\end{thebibliography}


	
	%\vfill
	

\end{document}